\documentclass[prl,twocolumn,showpacs,floatfix]{revtex4}

\usepackage{graphicx}
\usepackage{array}
\usepackage{amsmath,amsfonts,amssymb}
\usepackage[ansinew]{inputenc}

\newcommand{\figref}[1]{Fig.~\ref{#1}}
\newcommand{\uA}{$\mu$A}
\newcommand{\unit}[1]{\mathrm {\,#1}}
\newcommand{\ie}{{\em i.\,e.}}
\newcommand{\etal}{{\em et al.}}

\begin{document}

\title{Optical Probe of Quantum Shot Noise Reduction at a Single-Atom Contact}

\author{Natalia L. Schneider$^1$, Guillaume Schull$^2$, and Richard Berndt$^1$}

\affiliation{$^1$Institut für Experimentelle und Angewandte Physik, Christian-Albrechts-Universität zu Kiel, D-24098 Kiel, Germany\\
$^2$Institut de Physique et Chimie des Matériaux de Strasbourg, UMR 7504 (CNRS - Université de Strasbourg), 67034 Strasbourg, France\\}

\begin{abstract}
Visible and infra-red light emitted at a Ag-Ag(111) junction has been investigated from tunneling to single atom contact conditions with a scanning tunneling microscope.  
The light intensity varies in a highly nonlinear fashion with the conductance of the junction and exhibits a minimum at conductances close to the conductance quantum.
The data are interpreted in terms of current noise at optical frequencies, which is characteristic of partially open transport channels.
\end{abstract}


\maketitle

Owing to the particle nature of the electron, electrical current exhibits noise.
The noise spectral density provides information on the conduction process which is complementary to resistance data \cite{blanter}.
For quantum conductors with a conductance $G=G_0\sum_i T_i$ ($T_i$ is the transmission probability of a conducting quantum state $i$ and $G_0= 2 e^2/h$ is the conductance quantum) \cite{landauer,landauer2} the spectral density of the noise $P$ deviates from Schottky's classical result \cite{diode} $P_S = 2 e I$ ($I$ is the average current). 
$P$ is effectively reduced compared to $P_S$ as expressed by the Fano factor \cite{Lesov,Buttiker1990}
\begin{equation}
F = P / P_S = \sum_i T_i(1-T_i) / \sum_i T_i.
\label{ff}
\end{equation}
The noise amplitude is particularly large for partially open channels, $T_i \neq 1$, where electrons may be either reflected or transmitted, and a drastic reduction occurs when a transport channel is fully open $(T_i=1)$ and no reflection takes place.
At low frequencies ($\nu < 100 \unit{kHz}$), this effect has been evidenced for quantum point contacts (QPC) \cite{glattli,liyp} and atomic scale metallic contacts \cite{Brom1999}. 
Following a theoretical suggestion of Aguado and Kouwenhoven \cite{Aguado2000}, Ghz fluctuations of current in mesoscopic structures have been detected by investigating the transport properties of single \cite{Onac2006,Gustavsson2008} and double \cite{Khrapai2006,Gustavsson2007} quantum dots, superconducting mesoscopic photon detectors \cite{Deblock2003}, or by using low noise amplifiers followed by detectors \cite{Zakka-Bajjani2007}. 
In these experiments, the quantum detector is located close to the QPC and is sensitive to current shot noise.

Several mechanisms have been proposed to explain how the energy is transferred from the quantum conductor to the quantum detector \cite{Khrapai2006}. 
While a recent experiment by Gustavsson \etal\ indicates that the process is probably mediated by photons \cite{Gustavsson2008}, no direct observation was reported. 
To enable direct detection of photons, generating current noise in a quantum conductor at frequencies which correspond to visible light would be advantageous. 
Here, we present the first measurements of shot noise reduction at optical frequencies, which are 3 orders of magnitude higher than previously reported \cite{Gustavsson2008}.
To this end, we applied elevated voltages $V$ up to 2 V to Ag-Ag(111) junctions in a scanning tunneling microscope.
Light emitted at the junction was recorded in the far field with detection limits 240 THz $ < \nu < $ 750 THz, which were determined by localized plasmon resonances of the junction and detector response.
The conductance of Ag--Ag junctions can be controllably varied close to the contact point enabling significantly more detailed measurements than a previous experiment  with Au--Au junctions \cite{Schull2009}.
The intensity of the emitted light exhibits a clear minimum near $G=G_0$, confirming that the light is emitted by current fluctuations.
This is the first experimental demonstration of the role of current noise in STM-induced light emission. 
Moreover, in contrast to expectations   \cite{Zakka-Bajjani2007,Aguado2000,Onac2006,Gustavsson2007}, shot noise is observed up to frequencies exceeding the limit $\nu =  eV/h$.
This behavior is interpreted in terms of nonlinear electronic processes caused by the intense current passing the atomic size conductor.

The experiments were performed with an ultra high vacuum STM at low temperature (5.8 K).
Ag(111) surfaces as well as chemically etched W tips were cleaned by Ar$^+$ bombardment and annealing.
To increase the plasmon enhancement the W tips were coated with Ag by indenting them into the sample.
As a final step of preparation the tips were repeatedly approached to surface at a sample voltage $V = 1.3 \dots 2.0 \unit{V}$. 
This procedure was found to increase the stability of the tips for contact experiments.
The unusual currents $I$ (typically up to 200 \uA) and sample voltages $V$ (up to 2.0 V) used in the present experiments can lead to significant modifications of both tip and surface. 
Therefore, topographic images were recorded before and after the acquisition of spectra or the current was monitored during data acquisition to detect abrupt changes, which signal changes of the junction.
All data presented here were acquired without changes of the tip or sample.
Photons emitted at the tunneling junction were guided with an optical fiber to a grating spectrometer and a liquid nitrogen cooled CCD camera.
Spectra are not corrected for the wavelength dependent efficiency of the detection equipment, which has been reported in Ref.\ \onlinecite{hoffmann:305}.

\begin{figure}[htp]
\centering
\includegraphics[width=0.99\linewidth]{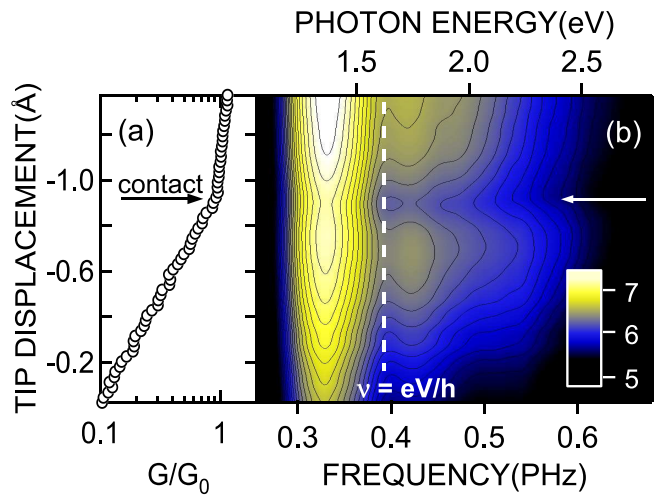}
\caption{(a) Conductance (in units of the conductance quantum $G_0=2e^2/h$) recorded during tip displacement $\Delta z$ from tunneling ($\Delta z = 0$) to contact. 
Sample voltage $V = 1.6\unit{V}$.
(b) Series of 60 luminescence spectra recorded simultaneously.
Intensities (counts per s and eV) are represented as false colors on a logarithmic scale.
Arrows indicate the formation of a single-atom contact.}  
\label{figure1}
\end{figure}

\figref{figure1} displays raw data on the light emission from a single atom contact on Ag(111).
During a stepwise approach of the STM tip the conductance (\figref{figure1}(a)) and spectra of the emitted light (\figref{figure1}(b)) were recorded.
The conductance trace exhibits a transition from tunneling to contact
(indicated by an arrow) at $G_c \approx 0.93 G_0,$ which is typical of single-atom Ag contacts \cite{nag_03,lli_05}.
We note that the smooth evolutions of the conductance and the spectra over the entire range of tip displacements provide additional evidence that no changes of the tip or the sample occurred during the measurement \cite{note}. 

The light emitted from the junction is composed of two components, similar to previous reports from Na monolayers  \cite{natrium} and Au(111) surfaces \cite{Schull2009}.
The low-energy emission at photon frequencies $\nu < e V / h$ is consistent with inelastic one-electron processes.
The high-frequency part of the spectra ($\nu > e V / h$, cf.\ dashed vertical line in \figref{figure1}b) is attributed to processes involving electrons which have been promoted to energies well above the Fermi level via electron-electron scattering.
For simplicity, we denote these components $1e$ and $2e$ light.

\begin{figure}[htp]
\centering
\includegraphics[width=0.99\linewidth]{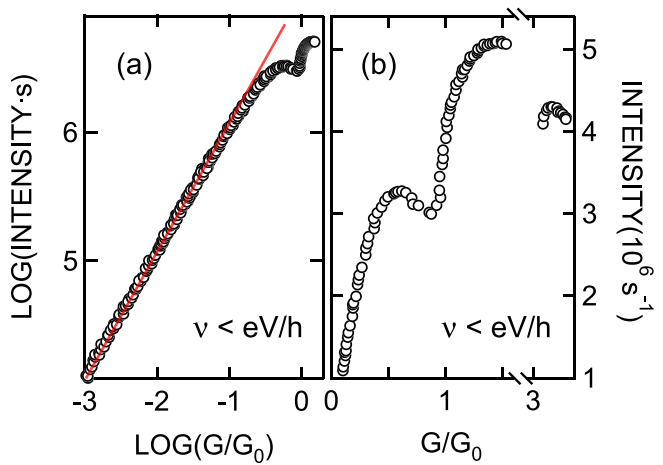}
\caption{ 
Intensity of $1e$ light vs.\ conductance.
Data recorded at $V=1.6 \unit{V}.$
The $1e$ intensity was integrated over $1.19 \unit{eV} < h \nu < 1.47 \unit{eV}.$
(a) Double-logarithmic scale covering conductances from $10^{-3}G_0$ to $1.6 G_0.$
For comparison, a straight line indicates a linear relation.
(b) Same data displayed on linear scales.
Occasionally, intensity data close to integer multiples of $G_0$ could be recorded and exhibited non-monotonous variation. Here, data for $G \gtrsim 3 G_0$ is included.}  
\label{figure2}
\end{figure}

The variations of the $1e$ and $2e$ light intensities with the displacement $\Delta z$ (and, hence, the conductance $G$) are qualitatively similar.
Given the additional complexity of $2e$ processes we now focus on an analysis of the $1e$ emission, which is well understood in the tunneling range \cite{PJ,PB}.
The $1e$ intensity as a function of the conductance is displayed in \figref{figure2}(a). 
Throughout the tunneling range, at conductances $G \lesssim 5\cdot10^{-2} G_0$, the intensity increases and is proportional to $G$ as expected \cite{PJ}.
Closer to $G=G_0$, however, clear deviations occur, which are more obvious on the linear scales of \figref{figure2}(b).
The intensity decreases despite the continued increase of the current 
and reaches a minimum at the point of contact formation (\figref{figure1}(b), arrow).
Beyond contact formation, another intensity rise occurs.
Similar behavior occurred near integer multiples of the quantum of conductance as illustrated
in \figref{figure2}(b) for $G \approx 3G_0$  \cite{note1}.

\begin{figure}[htp]
\centering
\includegraphics[width=0.99\linewidth]{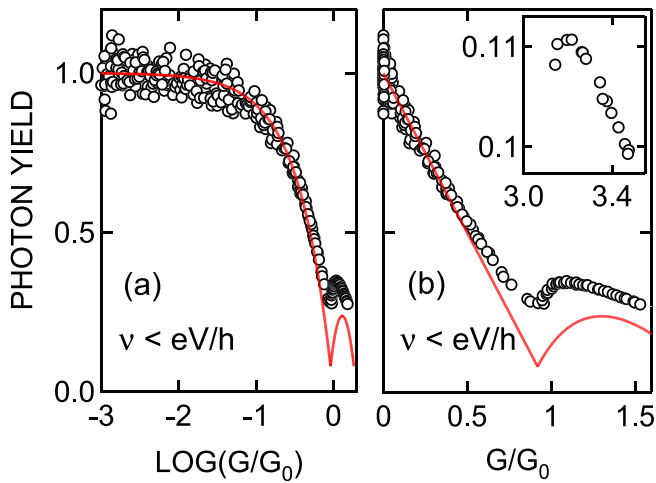}
\caption{
Yield of $1e$ photons (defined as $1e$ intensity divided by the current, normalized to 1 at low conductances) vs.\ conductance on (a) logarithmic and (b)   linear scales. 
Experimental data (dots) from \figref{figure1}b are compared to the Fano factor from Eq.\ \ref{ff}.
Inset to (b) shows data near $3 G_0$. A yield of 1 corresponds to an estimated quantum efficiency of $3\times10^{-6}$ photons per electron \cite{eff}.}  
\label{figure3}
\end{figure}

Light emission in the tunneling range, \ie\ at low conductance, is mediated by tip-induced plasmon modes (TIP modes), which are induced by the proximity of the STM tip to the sample \cite{PJ,IET}.
These modes, which can decay radiatively, enhance the intensity of the light emitted at TIP resonance frequencies, which, in the case of Ag--Ag junctions, enables light detection in the visible and near infra-red range. 
Except for a small red-shift ($\approx 10 \unit{nm}$) of the $1e$ peak \cite{aiz02} with tip approach, no major change of the spectral shape was observed, implying that the TIP modes remain largely unchanged over the
tip excursion of $\approx 0.14 \unit{nm}$ probed in Fig.\ \ref{figure1}.
The drastic 70\% decrease of the photon yield (\figref{figure3}) can hardly be explained by this red-shift.
However, the plasmon modes are driven by fluctuations of the tunneling current \cite{Rendell1978,Laks1979,PJ}.  
For this reason, the light intensity may be expected to be proportional to the high-frequency shot noise of the tunneling current. 
We propose the hypothesis that this linear relation remains valid close to and at contact.
The reduced light intensity near $G=G_0$ would then reflect the different noise characteristics at large conductances.
Indeed, model calculations as well as experiments have shown that the low-frequency shot noise of a quantum contact is reduced below its classical value as described by the Fano factor of Eq.\ \ref{ff} \cite{Lesov,Buttiker1990,glattli,liyp,Brom1999}. 

To test our hypothesis, \figref{figure3} shows a comparison of the yield of $1e$ light (circles), which we define as the $1e$ intensity per current, and the Fano factor (line).
The yield is normalized to 1 at low conductance values.
As a minimal model of the Fano factor, we used two conductance channels, similar to the approach of van den Brom and van Ruitenbeek \cite{Brom1999} (specifically, for $G \leqq 0.93 G_0$, we set $T_1= G/G_0$ and  $T_2= 0$ and, for $G > 0.93 G_0$,  $T_1= 0.93$ and  $T_2= G/G_0 - 0.93$).
\figref{figure3}(a) covers more than three decades of conductance on a logarithmic scale.
The experimental results in the tunneling range, 
which corresponds to the  limit of an almost closed conductance channel,
are consistent with the previous data reported by various researchers from various samples, where a linear relationship between intensity and current was observed.
Clear deviations from this linearity become obvious for $G/G_0 \gtrsim 0.2$, where a drastic reduction of the yield is observed.
The experimental data is described rather accurately by the Fano factor model.

\begin{figure}[htp]
\centering
\includegraphics[width=0.65\linewidth]{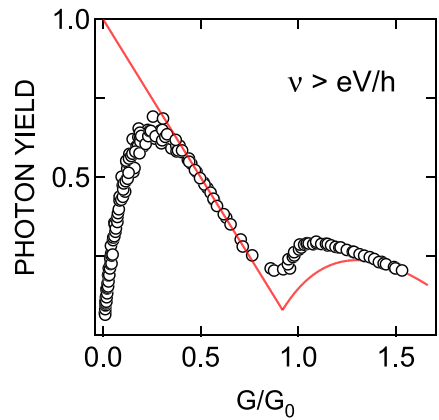}
\caption{Yield of $2e$ light ($2e$ intensity divided by current, arbitrarily normalized) vs.\ conductance. 
The intensity was integrated over $1.6 \unit{eV} < h \nu < 2.08 \unit{eV}.$
Experimental data (dots) from \figref{figure1}b are compared to the Fano factor from  \figref{figure3}. A yield of 1 corresponds to an estimated quantum efficiency of $3\times10^{-7}$ photons per electron \cite{eff}.}  
\label{figure4}
\end{figure}

The yield at conductances close to $G_0$ is shown in more detail in \figref{figure3}(b) on a linear scale.
It assumes a minimum at $G \approx 0.93 G_0$, \ie\ the conductance typically observed from a single atom Ag contact (cf.\ \figref{figure1}(a)).
At higher conductances, upon further approach of the STM tip, the yield recovers, and goes through a maximum followed by another reduction.
These variations are qualitatively reproduced by the Fano factor within the two-channel model. 
Similarly, the variation of the yield between 3 and 4 $G_0$ (\figref{figure3}(b) inset) is consistent with the Fano factor describing a 3 atom contact\cite{Brom1999}.
This lends strong support to the hypothesis that the $1e$ light intensity is directly related to the current noise.

The role of shot noise of quantum conductors as a source for the electromagnetic radiation has been theoretically discussed \cite{lelo,gavi,Aguado2000,bee,bee2} and 
indirect experimental evidence of the emission of light was found from the transport properties of various quantum detectors located in the near field of a noise source \cite{Onac2006,Gustavsson2008,Khrapai2006,Gustavsson2007,Deblock2003,Zakka-Bajjani2007}.
The present experiment is the first one to directly demonstrate that the current shot noise of a quantum point contact leads to the emission of sub-PHz radiation into the far field. 

Despite the similarities of the model and the data, there are deviations.  
Most notably, the intensity reduction at $G\approx 0.93 G_0$ is less complete than expected from the model.
This may partially be due to the presence of more than two open channels.
While previous results from Ag break junctions indicate that the contribution of additional channels is minor \cite{Ludoph2000}, the significantly higher bias voltages used in the present experiments are expected to modify the related transmission probabilities.
Heating of the electron gas is another likely explanation of the incomplete noise reduction.
Thermal fluctuations at a temperature $T$ lead to the following noise spectral density  \cite{Buttiker1992}
$$
P_T=2eVG_0 \coth\left(\frac{eV}{2kT}\right) \sum_i T_i(1-T_i)+4kT G_0 \sum_i T_i^2.
$$
Using this expression, the limited intensity reduction in the experiment can be fitted \cite{note2} assuming an electron temperature $T\approx 2000 \unit{K}$.

The yield of $2e$ light varies in an intriguing manner (\figref{figure4}).
While it is zero in the limit of vanishing conductance, an increase is observed up to $G \lesssim 0.25 G_0$.
At higher conductances, the experimental data approaches the behavior observed for $1e$ light, \ie, it is rather close to the Fano factor (solid line). 
As previously demonstrated $2e$ light may be attributed to inelastic transitions involving electrons which have been promoted to energies above the Fermi edge through electron-electron scattering \cite{natrium,Schull2009}.
We propose that the number of these electrons is low at low $G$ and effectively limits the yield.
At sufficiently high $G$, however, the yield tends to reflect current noise at frequencies $\nu > eV/h$ as described by the Fano factor.

In summary, the intensity of light emitted from single atom silver contacts exhibits a highly nonlinear variation with the conductance.
The intensity variation is interpreted in terms of a progressive reduction of high frequency shot noise.
An unexpected noise component at $\nu > eV/h$ is observed and interpreted in terms of non-linear electronic processes.
Finally, the results reveal a link between plasmon-mediated inelastic tunneling and current noise and raise intriguing questions into the interaction of electrons in single atom contacts with local optical modes.

We thank F. Charra, T. Frederiksen and P. Johansson, for discussions.
Financial support by the Innovationsfonds Schleswig-Holstein is gratefully acknowledged.


\begin{thebibliography}{32}
\bibitem[{not()}]{note}
\bibinfo{note}{When contacting a clean Ag(111) terrace no transfer of an atom
  from the tip to the surface was observed at the voltages used here, in
  contrast to the case of low bias voltages \cite{lli_05}. Spectroscopic
  results and conductance data were very similar for contacts to flat areas and
  adatoms.}
\bibitem[{not()}]{note1}
\bibinfo{note}{
The complex pattern of decreasing and increasing intensity is clear indication that effects such as shadowing
of the emission, changes of the emission direction close to contact or a voltage drop in the contact region can be excluded as possible origins of the intensity variations.
}
\bibitem[{not()}]{note2}
\bibinfo{note}{
The range of minimum photon yield, where the
Fano factor is minimal at $T=0$, was used for the fit.
}
\bibitem[{not()}]{eff}
\bibinfo{note}{ For an order-of-magnitude estimate of the absolute quantum efficiencies 
we considered the detected solid angle and assumed an isotropic emission.
The wavelength dependent efficiency of the detection setup was taken into account.}
\end{thebibliography}
\end{document}